\documentstyle[A4,11pt,psfig]{article}
\newcommand{\beq}{\begin{eqnarray}}
\newcommand{\eeq}{\end{eqnarray}}
\begin{document}
\title{On the derivation of the neutrino oscillation length formula.}
\author{J.-M. L\'evy \thanks{Laboratoire de Physique Nucl\'eaire et de Hautes Energies,
CNRS - IN2P3 - Universit\'es Paris VI et Paris VII, Paris.   \it Email: jmlevy@in2p3.fr}}
\pagenumbering{arabic}
\sloppy
\maketitle
\begin{abstract}
Forsaking the traditionnal hand-waving in the treatment of the motion, we show that the ultra-relativistic
approximation and the equality of kinematical variables are unnecessary ingredients in the derivation of the
oscillation length using plane waves, at least in a two flavor world.
It ensues that the formula is valid as it is in the non relativistic regime, provided one uses the correct
variable which is found to be momentum, not energy, and that the precise production kinematics is irrelevant.
Consequences for the more complete treatments are briefly evoked.
\end{abstract}
\newpage
\begin{center}
\end{center}
\par
The expression of the neutrino oscillation length is usually derived in the ultra-relativistic regime using
a superposition of plane waves supposed to have either equal momenta or equal energies. 
\\In the course of the derivation, the approximation $t \approx x$ is often made (see e.g. \cite{KAY})
to identify the oscillating pattern as a function of the distance from the production point. 
It is readily seen, however, that the correct result
obtains in such a treatment because the equal $p$'s or equal $E$'s hypothesis reduces  the time dependence or
the space dependence of the oscillation amplitude to an overall phase factor which disappears upon calculating
probabilities. \\
We will show that making the $t \approx x$ 'approximation' is unnecessary and can lead to senseless 
results if one uses the equal velocity hypothesis, which is admittedly neither better nor worse than
the above mentionned two other possibilities.\\

The more correct treatment given below hinges on a definition of  the (pseudo) 'center' of the would-be wave packet 
and shows that the usual oscillation length also obtains in the non relativistic case. 
In the end, it allows to get rid of {\bf any 
hypothesis} about the kinematics of the production process.\\

The meaning of these findings is, however, particulary clear in a simplified two-flavor world, where there is
but one oscillation length. The relevance of all this to a more realistic situation is briefly discussed.

\section{Three derivations}
Let us represent the neutrino born at space-time point $(0,0)$ in some charged current reaction involving charged 
lepton $l$ by
\beq |0,0> = |\nu_l> = \sum_h U_{l h}|h> \label{heav} \eeq where $|h>$ is a mass eigenstate with eigenvalue $m_h$ and
definite energy and momentum.\\
The fate of this neutrino is governed by the space-time translation operator\\ ${\cal U}=e^{-i(Ht-\vec{P}\cdot\vec{r})}$
and the problem amounts to correctly evaluate its action on (\ref{heav}):
\beq {\cal U}|0,0> = |x,t> =  \sum_h U_{l h} e^{-i(E_ht-p_hx)}|h> \label{act} \eeq
where we have assumed that the propagation  is along the $x$ axis. The precise values of $E_h$ and $p_h$ in
this formula depend on the production kinematics. In the case of a $\pi_{l2}$ decay for example, they are
fixed by the masses in the $\pi$ rest-frame and from there in the lab, once the $\pi$'s decay angle and velocity are given.
Projection of $|x,t>$ onto $|\nu_{l'}>$ yields then
\beq A(l \rightarrow l')(x,t) = \sum_h U^*_{l' h}U_{l h}e^{-i(E_ht-p_hx)} \label{amplit}\eeq
for the amplitude to detect a neutrino of flavor $l'$ at point $x$ and time $t$ relative to the
production point at $(0,0)$, granted that the neutrino interacts. \\
Henceforth, we shall assume a two flavor-two mass world, which simplifies the matter greatly.\\
The all-important object is then the one phase difference which appears upon squaring (\ref{amplit}):
\beq P(l \rightarrow l'\neq l)(x,t) = sin^2(2\theta) sin^2( \frac{\delta \phi}{2}) \eeq
where we have reverted to the simplified notation in use in the two flavor world and abbreviated:
$\delta \phi = \delta E t -\delta p x$\\
 
 In order to make contact with experiments which register the coordinates but not the time and because the
object of study should be more properly described by some more or less localized wave function, a connection between
$x$ and $t$ must be made at some point to describe what shall be considered as the motion of the center
of the wave packet. Also, the phase difference should be expressed in terms of the quantities which are really
at stake, {\it viz.} the masses and a single kinematical value representing the average energy or
momentum of the beam. To this effect, various supplementary hypotheses are added to the basic ingredient
represented by formula (\ref{act}) \\
None of these is necessary in the case at hand as we shall see now, 
provided the first problem is properly treated.
\subsection{Equal energies}
\label{eqe}
It is assumed here that the two massive components have the same energy and different momenta;
then $\delta \phi = -\delta p x$ and the oscillation pattern is described by:
$$O(x) = sin^2(\frac{\delta p x}{2})$$
Invoking a relativistic situation, people usually expand $p \approx E - \frac{m^2}{2E}$
hence $\delta \phi \approx \frac{\delta m^2}{2E}x$ and \footnote{$\delta$ is a true, signed, difference,
and $\Delta$ is an absolute value}
\beq O(x) \approx sin^2(\frac{\Delta m^2}{2E}x) \label{app1}\eeq
However, this is unnecessary since the exact relation: \beq\delta p^2 = \delta E^2 -\delta m^2 \label{core}\eeq
yields in this case: $$\delta p = -\frac{\delta m^2}{\Sigma p} = -\frac{\delta m^2}{2 \bar{p}}$$ with an
obvious definition for $\bar{p}$.\\
Consequently: \beq O(x) = sin^2(\frac{\Delta m^2}{4\bar{p}}x)& \rm{and}& L_{osc} = \frac{4\pi\bar{p}}{\Delta m^2} 
\label{exact}\eeq
Simple as is it, the meaning of this procedure is completely clear only in the case of two masses, 
since in the more general situation we could not 
introduce the third momentum into the definition of $\bar{p}$.

\subsection{Equal momenta} \label{eqp}
Here, $\delta \phi = \delta E t$ and using again a first order expansion, this becomes:
$\delta \phi \approx  \frac{\delta m^2}{2p}t$ after which the further 'relativistic approximations'
$t \rightarrow x$ and $p \rightarrow E$ allow to find consistency with the approximate result (\ref{app1}).\\
This again is unnecessary because ({\ref{core}) yields here $\delta E = \frac{\delta m^2}{\Sigma E}$
and, upon defining the velocity $v$ of the center of the would-be wave packet, one finds:
$$\delta \phi = \frac{\delta m^2}{\Sigma E} \frac{x}{v} = \frac{\delta m^2}{2p}x$$
provided $$v = \frac{2p}{E_1+E_2}$$ which shall be justified presently, but is seen to agree 
with the arithmetic
mean up to and including first degree terms in the small quantity $\frac{\Delta E}{\Sigma E}$\\

Hence \beq O(x) = sin^2(\frac{\Delta m^2}{4 p}x) & \rm{and} & L_{osc} = \frac{4\pi p}{\Delta m^2} \eeq
exactly as in (\ref{eqe}), granted the definition used for $v$, and with an analogous restriction since only
two energies must be considered in defining $v$.

\subsection{Equal velocities} \label{eqb}
In this case, $\delta \phi$ does not reduce to a single term and it is very important {\bf not} to approximate 
$t$ by $x$, for in so doing one would arrive at:
 $$\delta \phi = (\delta E - \delta p)x =  \delta m e^{-\eta}x$$ upon introducing $\eta = \tanh^{-1}(v)$\\

The oscillating pattern would be described by: 
\beq sin^2(\frac{\Delta m}{2}e^{-\eta}x)&
\rm{ and\ the\ oscillation\ length:\ } &L'_{osc}=\frac{2\pi e^{\eta}}{\Delta m}\eeq
However, since $$e^{\eta} = \frac{E_1+p_1}{m_1} = \frac{E_2+p_2}{m_2} \approx \frac{4p}{m_1+m_2}$$ this yields finally
\beq L'_{osc.}=\frac{8p\pi}{\Delta m^2} \label{false}\eeq {\it viz.} twice the usual value. \\

Confronted with this result, people have been tempted to think that the standard formula
is either false or does not apply in the case at hand. \\
A more carefull treatment of the motion of the would-be wave packet shows that this is not correct.
Indeed, if the hypothesis of equal velocities has any meaning, then the center of the wave packet
moves with {\bf that} velocity, not with velocity 1. Therefore, defining its position by 
$x = v t$ yields:
$$\delta \phi = \delta E (t-v x) = \delta m \gamma (1/v -v) x = \frac{\delta m}{v \gamma} x$$
\footnote{$\gamma = 1/\sqrt{1-v^2}$ as usual}
Now $$\frac{1}{v \gamma} = \frac{m_1}{p_1} = \frac{m_2}{p_2} = \frac{\Sigma m}{2 \bar{p}}$$
Hence $\delta \phi = \frac{\delta m^2}{2\bar{p}}x$ and the correct formula found in (\ref{eqe}) results.\\
The same restriction as before applies, since the third momentum cannot enter the definition of $\bar{p}$.\\

Clearly, replacing $1/v-v$ by $1-v$ (equivalent to $t \rightarrow x$) cannot be harmless; in (\ref{eqp}), $v$ is
only an overall factor, and replacing it by $1$ induces a relative error on the phase shift which goes to $0$ with 
$1-v$. Not so in the present case where the relative error is $1/(1+v)$ - hence the factor $2$ found above.
Stated differently, $v$ and $E$ were treated separetly in (\ref{eqp}) but here the connection between $m$, $v$ and
$E$ (or $p$) must be used.\\

\section{..and a fourth one.}
First observe that all three derivations above use exact relativistic kinematics but that none uses
any sort of 'ultra-relativistic approximation', especially not the ubiquitous but very unreasonable '$x\approx t$'.
A moment of reflexion reveals their common feature: in all three cases, the center of the would-be wave packet 
is endowed with the average momentum and energy of the components: $p^c=\bar{p}$, $E^c=\bar{E}$, and it is 
assumed to have velocity $v =\frac{p^c}{E^c}$. This is the real justification behind the definition of $v$ in (\ref{eqp}).\\
One is thus led to think that this is all that is needed to yield the well-known $L_{osc}$; indeed,
baring any {\it ad hoc} hypothesis on the production kinematics:
\beq 
\nonumber 
v = \frac{p_1+p_2}{E_1+E_2}  \Rightarrow & \delta \phi = \delta p x - \delta E t &= (\delta p - \delta E 
\frac{\Sigma E}{\Sigma p})x 
=\frac{(\delta p^2 -\delta E^2)x}{\Sigma p}\\ \nonumber && = \frac{-\delta m^2 x}{2\bar{p}} \eeq
which proves our point.
\section{Lessons}
\label{lesson}
The usefulness of all this is of course lessened by the well known shortcomings of the use of plane waves
for the purpose of describing neutrino oscillations (see e.g. \cite{R} for a list of these) ;
the necessity of using wave packets (see e.g.\cite{KRG})
or field theory (\cite{GMS},\cite{Car1}) has been the subject of a long and still ongoing debate and many sophisticated
treatments have appeared over the years. However, these more elaborate methods together with the inclusion
of the neutrino production and/or detection processes in the description of the phenomenon (\cite{R},\cite{C}) all 
result in
formulae which are subtended by the basic oscillation pattern described by (\ref{exact}), provided there 
exists a middle-zone where coherence is not lost but finite source 
length and momentum spread effects are negligible. 
In all cases, the same (vacuum) oscillation length obtains when and where 
resolution or decoherence do not blur the oscillations.\\ 

The above demonstration sheds some light on this robustness of the classical formula, by showing that none
of the extra hypotheses usually made is necessary, at least in the two flavor world 
where the oscillation length has its clearest meaning. Buried at the heart of the more sophisticated treatments 
is always some definition of the $x \leftrightarrow t$ relationship which avoids the hand-waving $x\approx t$.  \\

It is also seen that the relevant variable is the momentum,
not the energy, when the distinction applies; provided one uses this
variable, the standard result seems to follow also in the non relativistic regime. This might
be usefull if slow, Karmen-anomaly-like objects \cite{KAR} are confirmed;
consideration of non-relativistic effects in oscillations
due to such states have already appeared in the litterature (see \cite{AHL})\\

A note of caution is, however, in order: it is not entirely clear that an oscillation probability -and therefore,
an oscillation length- has a meaning 
in itself in the non-relativistic regime; the wave packet treatment (except in its most primitive form) and
the field theoretical approach both include the production and/or detection processes in an overall probability
calculation which generally factorizes in the relativistic regime, where all masses are small with respect to the
kinetic energy scale. 
It should be determined under what conditions this also applies in the non relativistic regime. Since oscillations can 
only occur 
in the case of nearly degenerate mass states, the production phase-space mass dependence should not be of 
concern,  but the detection reaction is a potential source of problem in that, e.g. the $\nu_{\tau}$ and 
possibly the $\nu_{\mu}$ component
can be inhibited for lack of energy; one can really appreciate at this point 
how much ill-defined are the so called 'weak eigenstates' (besides the fact that they are not eigenstates of
any operator which distinguishes them from the mass eigenstates!) 
\\Moreover, one must expect a larger yield of 'wrong' helicity when $\gamma \rightarrow 1$ and therefore an 
additional entanglement
 of 'flavor' with the other variables which might further preclude the definition of an 'oscillation probability' 
disconnected from the rest of the process \cite{Car}.

\section{Summary and conclusion}
Simple calculations show that none of the hypotheses usually employed in deriving
the neutrino oscillation length in the plane wave formalism is necessary and that the only requirement 
is a proper treatment of the motion of the would-be wave packet, at least in a two-mass world. 
This should apply to real life whenever the number of active mass states is reduced to two, in particular when
production phase-space is restricted or in case of degeneracy.\\

Although a plane wave treatment of neutrino oscillations is, admittedly, an over-simplification, we believe that
what has been done here has the merit of giving some clues in answering questions as to what are the 
conditions that should be met to allow observation of oscillations or that should be hypothetized in a sensible 
theoretical treatment. 
In particular, it casts some shadow on the relevance of the equal 
energies versus equal momenta arguments that have appeared in the litterature, given that 
the one important ingredient seems to be a suitably defined velocity for the
center of the wave packet representing the 
object under study.

\newpage
\end{document}